\documentclass[%
reprint,
groupedaddress,
showpacs,preprintnumbers,
 amsmath,amssymb,
 aps,
prb,
]{revtex4-1}

\usepackage{graphicx}
\usepackage{dcolumn}
\usepackage{bm}

\usepackage{txfonts}
\usepackage{color}

\begin{document}

\preprint{APS/123-QED}

\title{High-field magnetization and magnetic phase transition in CeOs$_2$Al$_{10}$}

\author{Akihiro Kondo$^1$}
 \email{kondo@issp.u-tokyo.ac.jp}
\author{Junfeng Wang$^{1,2}$}%
\author{Koichi  Kindo$^1$}
\author{Yuta Ogane$^3$}
\author{Yukihiro Kawamura$^3$}
\author{Sakiyo Tanimoto$^3$}
\author{Takashi Nishioka$^3$}
\author{Daiki Tanaka$^4$}
\author{Hiroshi Tanida$^4$}
\author{Masafumi Sera$^4$}
\affiliation{%
$^1$Institute for Solid State Physics, University of Tokyo, Kashiwa, Chiba 277-8581, Japan
}%
\affiliation{%
$^2$Wuhan High Magnetic Field Center, Huazhong University of Science and Technology, Wuhan 430074, China
}%
\affiliation{
$^3$Graduate school of Integrated Arts and Science, Kochi University, Kochi 780-8520, Japan
}%
\affiliation{%
$^4$ Department of Quantum Matter, AdSM, Hiroshima University, Higashi-hiroshima 739-8530, Japan
}%


\date{\today}

\begin{abstract}
We have studied the magnetization of CeOs$_2$Al$_{10}$ in high magnetic fields up to 55 T for $H\parallel a$ and constructed the magnetic phase diagram for $H\parallel a$. 
The magnetization curve shows a concave $H$ dependence below $T_{\rm max}\sim$40 K which is higher than the transition temperature $T_0\sim$29K. 
The magnetic susceptibility along the $a$-axis , $\chi_a$ shows a smooth and continuous decrease down to $\sim$20 K below $T_{\rm max}\sim$40 K without showing an anomaly at $T_0$.
From these two results, a Kondo singlet is formed below $T_{\rm max}$ and coexists with the antiferro magnetic order below $T_0$.  
We also propose that the larger suppression of the spin degrees of freedom along the $a$-axis than along the $c$-axis below $T_{\rm max}$ is associated with the origin of the antiferro magnetic component.

\end{abstract}

\maketitle

Ce$T$$_2$Al$_{10}$($T$=Ru, Os, Fe) have recently attracted much attention because of their unusual physical properties.\cite{Stry,Muro0,Nishi,Matsu,Tany1,Han, mSR,Tany2,Kon,ND,Kha,Tany3,Kon2,Muro, Adro,Migno}
These compounds belong to the category of the Kondo semiconductor.
The most characteristic feature in this system is that CeRu$_2$Al$_{10}$ and CeOs$_2$Al$_{10}$ exhibit an unusual long range order (LRO).
These two compounds exhibit a large anisotropy of the magnetic susceptibility, $\chi$; $\chi_a>\chi_c>\chi_b$.  
 Here, $\chi_a$ etc. indicates $\chi$ along the $a$-axis, etc. 
 From their temperature dependences, these two compounds are categorized into roughly localized system. 
The LRO temperature of $T_0\sim$30 K is extremely high in comparison with $T_{\rm N}$ of other $R$$T_2$Al$_{10}$($R$=rare-earth element).  
$T_{\rm N}$ is equal to 16.5 K even in GdRu$_2$Al$_{10}$.
Among Ce$T$$_2$Al$_{10}$ compounds, CeRu$_2$Al$_{10}$ is most intensively studied.
At the early stage, Tanida $et$ $al$. proposed a singlet ground state from the following results. (1) A large magnetic entropy released at $T_0$, (2) A decrease in the magnetic susceptibility along all the crystal axes below $T_0$, (3) A positive pressure effect on $T_0$ in addition to the activation-type $T$ dependence of the physical properties.\cite{Tany1}
We confirmed the magnetic origin from the high field magnetization measurement of CeRu$_2$Al$_{10}$ and obtained the magnetic phase diagram.\cite{Kon}
Although a spin gap with an excitation energy of 8 meV, consistent with the singlet ground state, was observed by inelastic neutron scattering,\cite{ND} soon after, the magnetic ordering was reported.\cite{Kha}
The antiferro magnetic (AFM) order is characterized by an AFM moment ($M_{\rm AF}$) of $\sim$0.34$\mu_{\rm B}/$Ce parallel to the $c$-axis.\cite{Kha}
Although it is now confirmed that the magnetic ordering with $M_{\rm AF}\parallel c$ is realized below $T_0$, the magnetic ordered state is definitely not a simple one and there exist many unusual properties;  a very high $T_0$ despite a small ordered moment, no metamagnetic transition despite the existence of a large spin gap without a low energy magnetic excitation, the magnetic order with $M_{\rm AF}\parallel c$ even though $\chi_a$ is much larger than $\chi_c$ in the paramagnetic region.

CeOs$_2$Al$_{10}$ exhibits physical properties similar to those of CeRu$_2$Al$_{10}$ and the same type of magnetic order is expected to be realized below $T_0$. 
On the other hand, there exist the following differences.\cite{Nishi,Muro,Adro}
$T_0$ is equal to 28.7 K which is a little higher than that of CeRu$_2$Al$_{10}$, the magnetic entropy  released at $T_0$ and the magnitude of the ordered moment are smaller than those in CeRu$_2$Al$_{10}$, $\chi_a$ and $\chi_c$ in the paramagnetic region exhibit a broad maximum at $T_{\rm max}\sim$40 K in CeOs$_2$Al$_{10}$, the spin gap energy is 11 meV\cite{Adro} which is higher than 8 meV in CeRu$_2$Al$_{10}$.\cite{ND} 
By applying pressure to CeRu$_2$Al$_{10}$, the physical properties are changed from a nearly localized regime to valence fluctuation regime as in CeFe$_2$Al$_{10}$.\cite{Nishi}
CeOs$_2$Al$_{10}$ is situated just between these two compounds limits.  

To clarify the origin of the above differences between the two compounds, we measured the high field magnetization of CeOs$_2$Al$_{10}$. 


The single crystals of Ce$T_2$Al$_{10}$ ($T$ = Ru, Os) used in the present study were prepared by the Al self-flux method.
Pulsed magnetic fields up to 55 T were generated with a duration of 36 ms using non destructive magnets. 
Magnetization ($M$) was measured by the induction method using a standard pick-up coil in magnetic fields along the $a$-axis.  
The pressure effect of $M$ for $H\parallel a$ was also measured using a piston cylinder type pressure cell.

\begin{figure}[tb]
\begin{center}
\includegraphics[width=.88\linewidth]{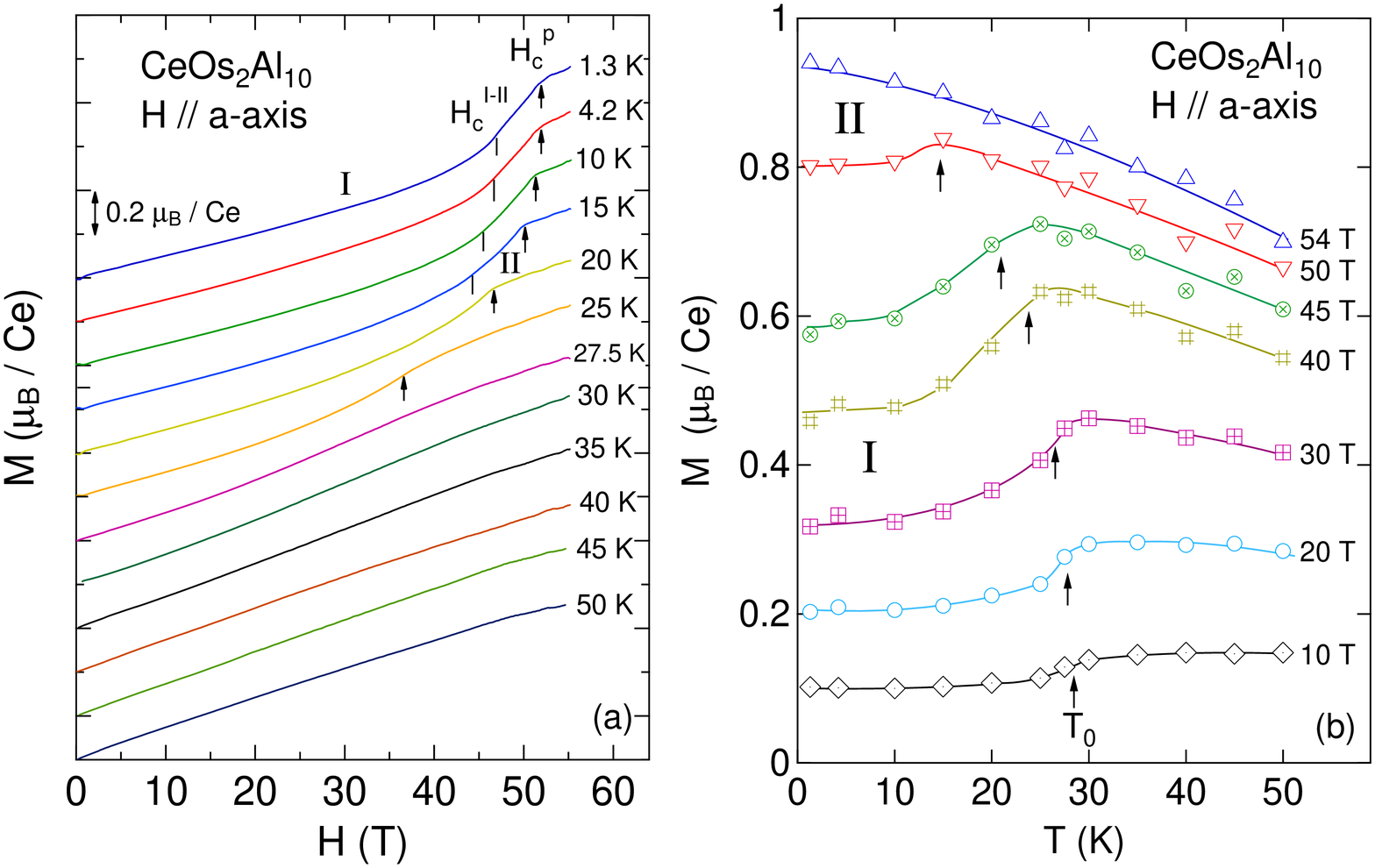}
\caption{(Color online) (a) Magnetic field dependence of the magnetization of CeOs$_2$Al$_{10}$ under various temperatures for $H\parallel a$-axis. The origin of the vertical axis of each curve is shifted.  
The ordered state is divided into two phases I and II. 
(b) Temperature dependence of the magnetization under various magnetic fields for $H\parallel a$-axis.
They are plotted by using the results in Fig.1(a).
The arrow indicates the transition temperature $T_0$, which is cited from the magnetic phase diagram shown in Fig. 2. }
\label{Fig1}
\end{center}
\end{figure}

Figures 1(a) and (b) represent the $M$-$H$ curves of CeOs$_2$Al$_{10}$ under various temperatures and the $T$ dependences of $M$ under various magnetic fields for $H\parallel a$, respectively.
The latter are obtained by using the results in Fig.1(a).
At $T$=1.3K, $M$ shows an $H$-linear increase up to $\sim$30 T and above $\sim$30 T, a concave $H$ dependence is observed.
After showing an anomaly at $H_{\rm c}^{\rm I-II}$=47 T, $M$ increases steeply and shows a saturated behavior at $H_{\rm c}^{p}$=52 T.
$M$ at $H_{\rm c}^p$ is 0.95 $\mu_{\rm B}/$Ce which is smaller than 1.3 $\mu_{\rm B}/$Ce in CeRu$_2$Al$_{10}$.\cite{Kon} 
Above 20 K, an anomaly at $H_{\rm c}^{\rm I-II}$ could not be recognized in the present experiments.
We note that although $M$-$H$ curves in the paramagnetic region above 40 K are normal, those at $T$=30 K and 35 K show a concave $H$ dependence even in the paramagnetic region,  which is shown in Fig. 3 (a) in an expanded scale. 
 Such an $H$ dependence is not observed in CeRu$_2$Al$_{10}$ above $T_0$.\cite{Kon}
As will be mentioned later, this different behavior of $M$-$H$ above $T_0$ between two compounds is observed as a different $T$ dependence of $\chi_a$ above $T_0$. 
The $T$ dependences of $M$ in magnetic fields between 10 T and 45 T are plotted on Fig. 1(b) and show a broad maximum several Kelvin above $T_0$ as was observed in the $\chi$-$T$ curve measured at low magnetic fields.
No clear anomaly is seen at $T_0$. 
At $H$=50 T, a maximum is seen at $T_0$=16 K and the $T$ dependence of $M$ below $T_{\rm N}$ is weak, different from a continuous decrease observed below $H$=45 T.
The $T$ dependence at $H$=54 T is that observed in a usual magnetic compound in the paramagnetic region.
Thus, the $H$ region above $H_{\rm c}^p$ is considered to be paramagnetic. 
This is also supported by the results of Ce$_x$La$_{1-x}$Ru$_2$Al$_{10}$.\cite{Tany1,Kon} 
$T_0$ and $H_{\rm c}^p$ is reduced roughly in proportional to La concentration toward $x\sim0.5$ and disappear.\cite{Tany1}

The magnetic phase diagram of CeOs$_2$Al$_{10}$ for $H\parallel a$-axis obtained from the results in Fig.1(a) is shown in Fig.2.
The inset shows the $M$-$H$ curve and $H$ dependence of d$M$/d$H$ at $T$=1.3 K.
Two anomalies are clearly seen at $H_{\rm c}^{\rm I-II}$=47 T and $H_{\rm c}^p$=52 T.
The magnetic phase diagram is similar to that of CeRu$_2$Al$_{10}$.\cite{Kon}
Hereafter, we call two LRO phases below and above $H_{\rm c}^{\rm I-II}$ as phase I and II, respectively. 
The region of phase II is narrower than that in CeRu$_2$Al$_{10}$.\cite{Kon}

\begin{figure}[tb]
\begin{center}
\includegraphics[width=.6\linewidth]{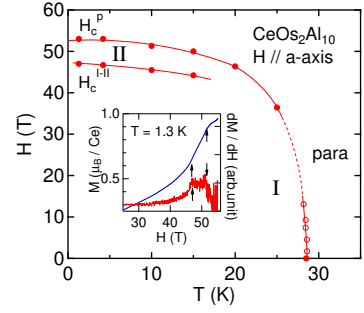}
\caption{(Color online) Magnetic phase diagram of CeOs$_2$Al$_{10}$ for $H\parallel a$-axis.  The open red circles below 14 T are cited from ref. 13.   The inset shows the magnetization curve (blue line) and derivative d$M$/d$H$ (red line) at $T$ = 1.3 K . }
\label{Fig2}
\end{center}
\end{figure}

Figures 3(a) and 3(b) represent the $M$-$H$ curves of CeOs$_2$Al$_{10}$ and CeRu$_2$Al$_{10}$ for $H\parallel a$ at $T$= 1.3 K, respectively.
In Fig.3(a), those at 30 K and 35 K are also shown and in Fig.3(b), that for $H\parallel c$ is also shown.
While both compounds show the concave $M$-$H$ curve for $H\parallel a$, a concave $H$ dependence is much more pronounced in CeOs$_2$Al$_{10}$ than in CeRu$_2$Al$_{10}$. 
The dashed straight lines highlight the difference of the degree of the concave $H$ dependence in two compounds. 
See also caption in Fig. 3. 
The concave $H$ dependence is also seen at 30 K and 35 K which is above $T_0$, although such an $H$ dependence is not seen in CeRu$_2$Al$_{10}$ above $T_0$.\cite{Kon,Kon2} 
Although the $M$-$H$ curve for $H\parallel c$ is not measured in CeOs$_2$Al$_{10}$, it is expected to be similar to that in CeRu$_2$Al$_{10}$ from the small magnitude of the magnetization, apart from a small anomaly at $H^*$ = 4 T and 6 T in CeRu$_2$Al$_{10}$ and CeOs$_2$Al$_{10}$, respectively.\cite{Tany3,Kon2,Muro} 
Also for $H\parallel b$, an $H$-linear $M$-$H$ curve is expected.
The concave $M$-$H$ curve is the characteristic feature for $H\parallel a$ in these two compounds.

Figure 4 represents the $T$ dependence of $\chi$ of CeRu$_2$Al$_{10}$ and CeOs$_2$Al$_{10}$ along the $a$-, $b$- and $c$-axes.  
The results show a large anisotropic behavior, $\chi_a>\chi_c>\chi_b$ in all the $T$ region.  At high temperatures, $\chi_c$ and $\chi_b$ show a similar $T$ dependence in both compounds.  
Along the $a$-axis, a Curie-Weiss behavior is more pronounced in CeRu$_2$Al$_{10}$ than in CeOs$_2$Al$_{10}$.  
The magnitude of $\chi_a$ in CeOs$_2$Al$_{10}$ at $\sim T_0$ is much smaller than that in CeRu$_2$Al$_{10}$.  
These indicate that Ce ion in CeOs$_2$Al$_{10}$ is closer to the valence fluctuation regime than in CeRu$_2$Al$_{10}$. 
A clear broad maximum is seen at $\sim$ 40 K along the $a$- and $c$-axes in CeOs$_2$Al$_{10}$.  
The $T$ dependence of $\chi_a$ at $\sim T_0$ is very different between two compounds.  
In CeRu$_2$Al$_{10}$, a clear anomaly is seen at $T_0$ and a rapid decrease below $T_0$.  
On the other hand, in CeOs$_2$Al$_{10}$, a continuous decrease is seen down to $\sim$20 K without showing an anomaly at $T_0$. 
The dotted lines for $\chi_a$ are drawn assuming that the paramagnetic state continues to exist down to $T$=0 K.  
Their extrapolation to $T$= 0 K is determined from the slope of the dotted straight lines in the $M$-$H$ curves in Fig.3.  
Namely, the dotted lines in Fig.3 corresponds to the $T$ dependence of $\chi_a$ just above $H_c^p$ , where the paramagnetic state is realized.   
We note that the size of the area below the dotted line above the experimental result in Fig.4 seems to correspond to that in the $M$-$H$ curve in Fig.3 in both compounds.

\begin{figure}[tb]
\begin{center}
\includegraphics[width=.83\linewidth]{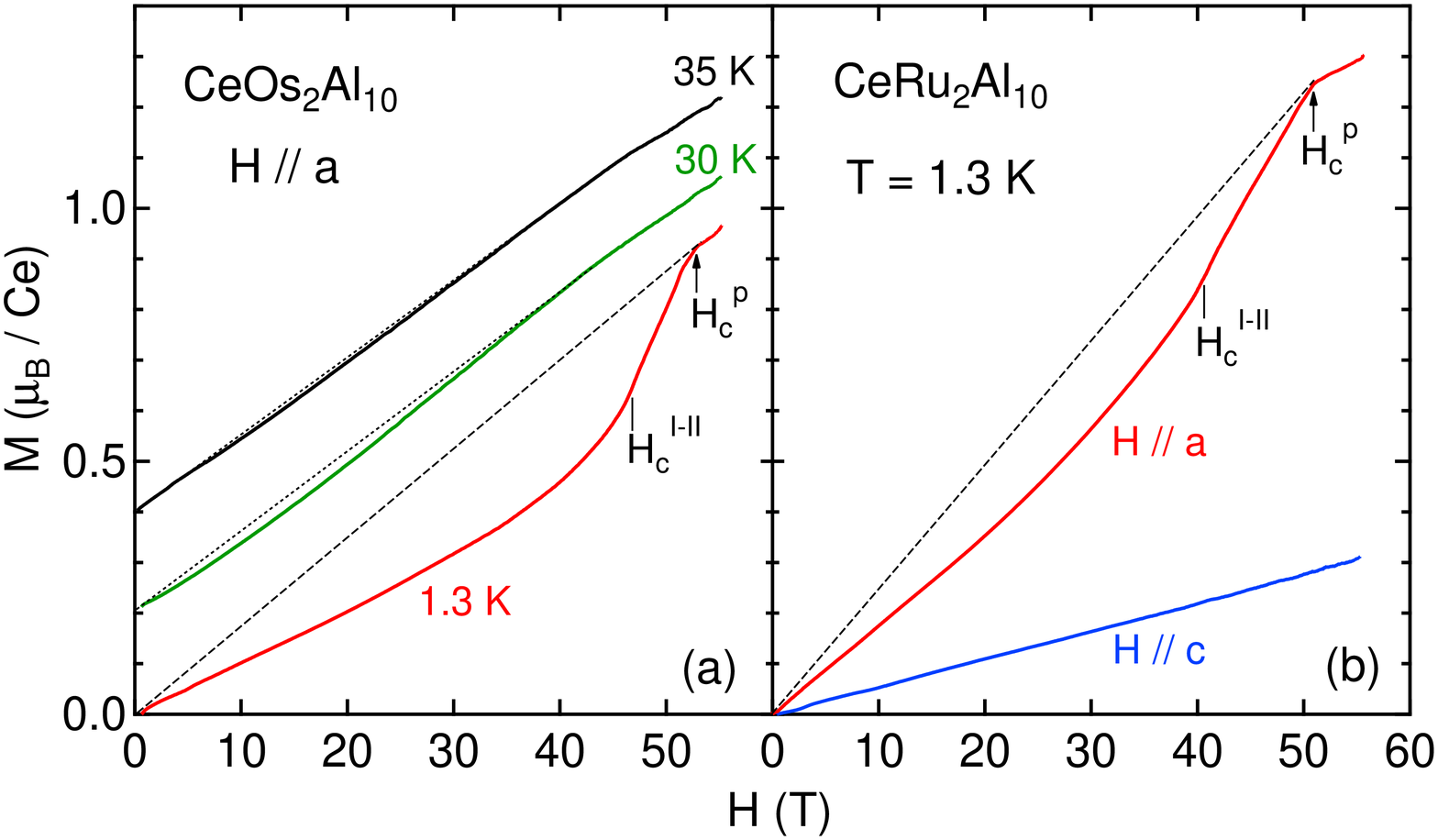}
\caption{(Color online) (a) Magnetic field dependence of the magnetization of CeOs$_2$Al$_{10}$ at $T$ = 1.3 K for $H\parallel a$-axis. Those at $T$ = 30 K and 35 K are also shown. The dotted lines for the results at $T$=30 K and 35 K are drawn only to see the concave $H$ dependence in these results. (b) Magnetization curve of CeRu$_2$Al$_{10}$ at $T$ = 1.3 K for $H\parallel a$-axis (red line) and $c$-axis (blue line). The data are cited from refs. 9 and 14. The dashed straight lines in Fig. 3 (a) and (b) are drawn by connecting the origin at $H$ = 0 and $M$ at $H_c^p$. }
\label{Fig3}
\end{center}
\end{figure}

Now, we discuss the relation between the large spin gap ($\Delta_{\rm SG}$) below $T_0$ observed in inelastic neutron scattering and the present results of magnetization, because these two are expected to be associated with each other closely.  
The relation between $\Delta_{\rm SG}$ and the Kondo temperature, $T_{\rm K}$ = 3$T_{\rm max}$ was discussed in ref. 12.  
The origin of $\Delta_{\rm SG}$ is difficult to be ascribed to an anisotropy gap of a magnon branch for the following reasons.  
The anisotropy of $\chi$  in CeOs$_2$Al$_{10}$ is smaller than that in CeRu$_2$Al$_{10}$ at low temperatures.   
The magnitude of the magnetic moment at $H_c^p$ and the ordered moment is smaller in CeOs$_2$Al$_{10}$ than in CeRu$_2$Al$_{10}$.\cite{Kha, Adro,Migno}  
These suggest that CeOs$_2$Al$_{10}$  is closer to the valence fluctuation regime than CeRu$_2$Al$_{10}$. 
This is also supported by the pressure effect of the electrical resistivity of CeRu$_2$Al$_{10}$.\cite{Nishi}    
These strongly suggest that the magnetic anisotropy in the ordered state is smaller in CeOs$_2$Al$_{10}$ than in CeRu$_2$Al$_{10}$.
This leads to a smaller magnon gap in CeOs$_2$Al$_{10}$ than in CeRu$_2$Al$_{10}$.  
However, experimentally a larger $\Delta_{\rm SG}$ is seen in CeOs$_2$Al$_{10}$ than in CeRu$_2$Al$_{10}$.  
Considering the fact that CeOs$_2$Al$_{10}$ is closer to the valence fluctuation regime and $T_0$ is higher in CeOs$_2$Al$_{10}$ than in CeRu$_2$Al$_{10}$, the hybridization between conduction band and nearly localized 4$f$ shell ($c$-$f$ hybridization) is expected to be associated with the origin of the large $\Delta_{\rm SG}$.   
By applying pressure to CeRu$_2$Al$_{10}$, $T_0$ increases and the physical properties become closer to those in CeOs$_2$Al$_{10}$.\cite{Nishi}
This indicates that the Ce-Ce interaction is enhanced by increasing the $c$-$f$ hybridization.  
This is different from many Ce compounds where the transition temperature is suppressed by pressure as a result of the suppression of the RKKY interaction by the Kondo effect.

\begin{figure}[tb]
\begin{center}
\includegraphics[width=.65\linewidth]{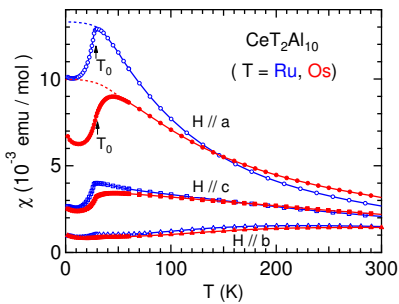}
\caption{(Color online) Temperature dependence of the magnetic susceptibility of CeRu$_2$Al$_{10}$ and CeOs$_2$Al$_{10}$ along the $a$-, $b$- and $c$-axes measured at 1 T. 
As for the dotted lines in $\chi_a$-$T$, please see the text in details. 
}
\label{Fig4}
\end{center}
\end{figure}

Based on these considerations, we discuss the magnetization of Ce$T_2$Al$_{10}$.  
If the AFM order with $M_{\rm AF}\parallel c$ were realized at $H$ = 0, a spin canting magnetization process would occur for $H\parallel a$ up to $H_c^p$.  
The $M$-$H$ curve of the present compounds show a pronounced concave $H$ dependence above $\sim 30$ T and another phase transition at $H_c^{\rm I-II}$ appears little below $H_c^p$. 
 Such an $M$-$H$ curve for $H\parallel a$ is difficult to explain by a spin canting magnetization process. 
In order to explain the concave $M$-$H$ curve in a spin canting magnetization process, the magnitude of the magnetic moment should increase accompanied with a decrease of a spin canting angle. 
The $T$ dependence of $\chi_a$, which is expected for $\chi_\perp$, is not seen in phase I but only seen in phase II. 
Such a spin canting magnetization process seems unphysical. 
Then, we expect that the origin of the concave $M$-$H$ curve is ascribed to the existence of the large spin gap.  
First, we discuss the $T$ dependence of $\chi$ below and above $T_0$.  
$\chi_a$ and also $\chi_c$ show a broad maximum at $T_{\rm max}\sim$ 40 K.  
The decrease of $\chi$ below $T_{\rm max}$ indicates that the spin degrees of freedom is suppressed with decreasing temperature even in the paramagnetic region.  
As its origin, the singlet formation could be considered. 
Considering that a broad maximum is seen in $\chi_a$ and $\chi_c$, the $c$-$f$ hybridization in the $ac$-plane may play an important role below $T_{\rm max}$.  
This is consistent with our previous pointing out that this system is viewed as the $ac$-plane two dimensional system stacked along the $b$-axis.\cite{Tany2}
The decrease of $\chi_a$ and $\chi_c$ below $T_{\rm max}$ in CeOs$_2$Al$_{10}$ is expected to be associated with the appearance of a large spin gap in the paramagnetic region. 
We expect that the spin gap observed in inelastic neutron scattering continues to exist up to $T_{\rm max}$.  
The ratio of $T_0$= 27.3 K in CeRu$_2$Al$_{10}$ to $T_{\rm max}\sim$ 40 K in CeOs$_2$Al$_{10}$ shows a good correspondence to the ratio of $\Delta_{\rm SG}$=8 meV in the former to 11meV in the latter.\cite{ND, Adro}  
In ref.12 on CeOs$_2$Al$_{10}$, it is noted the possibility that the inelastic peak exists at $T_0$.  
Thus, we conjecture that in CeOs$_2$Al$_{10}$, a spin gap appears below $T_{\rm max}$.  
This should be examined by the inelastic neutron scattering experiments. 
The concave $M$-$H$ curve for $H\parallel a$ remains even in the paramagnetic region up to $T_{\rm max}$.  
This comes from the same origin as a decrease of $\chi_a$ below $T_{\rm max}$.  
The $T$ dependence of $\chi_a$ of CeRu$_2$Al$_{10}$ is very different from that of CeOs$_2$Al$_{10}$.  
We measured the pressure effect of $M$ of CeRu$_2$Al$_{10}$ for $H\parallel a$. 
When the pressure is applied to CeRu$_2$Al$_{10}$, the magnitude of $M$ is suppressed and becomes to show a broad maximum above $T_0$.  
The magnetic properties of CeRu$_2$Al$_{10}$ become closer to those of CeOs$_2$Al$_{10}$ in the same way as in the pressure effect of the electrical resistivity.\cite{Nishi}   
As the $c$-$f$ hybridization is enhanced by applying pressure, the different behavior of $\chi_a$ around $T_0$ between CeOs$_2$Al$_{10}$ and CeRu$_2$Al$_{10}$ could be understood as a result of a different magnitude of the $c$-$f$ hybridization. 
As the $c$-$f$ hybridization should be associated with the singlet formation, the above mentioned singlet is considered to be a Kondo singlet as discussed in ref. 12. 
Considering that $\chi_a$ shows a continuous decrease without showing an anomaly at $T_0$, the decrease of $\chi_a$ below $T_0$ is also dominated by the same mechanism as that above $T_0$ below $T_{\rm max}$.  
From these results, we propose that the Kondo singlet accompanied with a spin gap begins to be formed below $T_{\rm max}$ and coexists with $M_{\rm AF}$ component parallel to the $c$-axis in the ordered state.  
As for $\chi_c$, although a broad maximum is seen at $T_{\rm max}$, a rapid decrease is observed below $T_0$ different from no anomaly in $\chi_a$.  
This rapid decrease of $\chi_c$ below $T_0$ may originate from $M_{\rm AF}\parallel c$ below $T_0$.
Now, as it is clear that the origin of the concave $M$-$H$ curve for $H\parallel a$ below $T_0$ is the same as that of the decrease of $\chi_a$ below $T_{\rm max}$, we could ascribe the former to the existence of a Kondo singlet in the ordered state.  
Namely, the destruction of a Kondo singlet by the magnetic field induces the concave $H$ dependence of $M$.  
The degree of the concave $H$ dependence is much more pronounced in CeOs$_2$Al$_{10}$ than in CeRu$_2$Al$_{10}$.  
This may indicate that the weight of the singlet component in the ordered state is larger in the former than in the latter.

Next, we discuss the reason why $M_{\rm AF}\parallel c$ is realized in the ordered state despite a large anisotropy of $\chi$, $\chi_a>\chi_c>\chi_b$,  in the paramagnetic region.  
The fact that $M_{\rm AF}\parallel c$ means that the anisotropic exchange interaction should exist and that along the $c$-axis, $J_{\rm ex}^c$ is largest. 
However, $J_{\rm ex}^c$ is considered to be not strong enough to fix $M_{\rm AF}$ to the $c$-axis tightly. 
This comes from the $\theta$ dependence of $M$ of CeRu$_2$Al$_{10}$ in the $ab$, $bc$ and $ca$ planes at constant temperatures below and above $T_0$.  
Here, $\theta$ is the angle between the applied $H$ direction and the crystal axis in each plane.  
We observed the smooth $\theta$ dependence of $M$ following $\cos\theta$ in all the planes.  
The $\theta$ dependence below $T_0$ is very similar to that above $T_0$.  
This indicates that the anisotropic exchange interaction which forces $M_{\rm AF}$ along the $c$-axis in the ordered state is small. 
Then, we propose the following explanation.  
In CeOs$_2$Al$_{10}$, the magnitude of the reduction of $\chi$ down to $T_0$ below $T_{\rm max}$ is $\sim$ 15 \% and $\sim$ 5 \% for $H\parallel a$ and $H\parallel c$, respectively.  
This means that the suppression of the spin degrees of freedom in the paramagnetic region below $T_{\rm max}$ is larger along the $a$-axis than along the $c$-axis. 
By considering that the suppression of the spin degrees of freedom comes from the $c$-$f$ hybridization, the larger hybridization along the $a$-axis could make a magnetization easy axis along the $c$-axis in place of the $a$-axis expected from the largest magnitude of $\chi_a$ in the paramagnetic region.

Next, we comment the possible origin of the large anisotropic magnetic susceptibility of Ce$T_2$Al$_{10}$ in the paramagnetic region.  
$\chi$ of Ce$T_2$Al$_{10}$ with a large anisotropy indicates that the crystalline electric field (CEF) splitting is large. 
On the other hand, our recent results of the magnetization and thermal expansion of Nd$T_2$Al$_{10}$ ($T$=Ru, Os) whose $T_{\rm N}$ is 2.4 K and 2.2 K in NdRu$_2$Al$_{10}$ and NdOs$_2$Al$_{10}$, respectively indicate that the CEF splitting in Nd$T_2$Al$_{10}$ is small.\cite{Tanimoto}
The Curie-Weiss law with $\mu_{\rm eff}\sim$3.65$\mu_{\rm B}/$Nd which is close to that of a free Nd ion is observed above $\sim$100 K along all the crystal axes in both compounds. 
The distortion below $T_{\rm N}$ is as small as $10^{-6}$.
These results indicate that the coupling between the lattice and the orbital moment of Nd ion is very small and the Nd ion behaves as a rather free ion, which may come from the caged structure surrounded by Al and $T$ ions.  
From these results, the Coulomb interaction as the origin of the CEF splitting is expected to be small also in Ce$T_2$Al$_{10}$, which contradicts with the large anisotropic magnetic susceptibility. 
Thus, we expect that in Ce$T_2$Al$_{10}$, a $c$-$f$ hybridization effect is associated with the large CEF splitting.  
The same $c$-$f$ hybridization effect may be also associated with the opening of a spin gap below $T_{\rm max}$ and the phase transition at $T_0$. 
Thus, the long range order in Ce$T_2$Al$_{10}$ seems to be dominated by the unusual Ce-Ce interaction.

Finally, we comment the CDW transition proposed by Kimura $et\ al$. quite recently. \cite{Kimura} 
Although the optical conductivity clearly shows that the electronic structure along the $b$-axis is modulated, the CDW transition along the $b$-axis seems difficult from the following results in CeRu$_2$Al$_{10}$.\cite{Nishi,Tany2} 
The electrical resistivity shows a clear enhancement below $T_0$ along all the crystal axes. 
The largest enhancement is seen not along the $b$-axis but along the $c$-axis. 
The Hall resistivity measured in $H\parallel a$ where the Hall current is in the $bc$ plane does not show an anomaly at $T_0$.\cite{Tany2} 
Although these are the results in CeRu$_2$Al$_{10}$, considering that the same type of LRO is strongly expected in these two compounds, the above results are difficult to understand by the CDW transition along the $b$-axis. 
The fact that the temperature where a charge gap along the $b$-axis begins to open coincides with $T_{\rm max}$ of $\chi_a$ and $\chi_c$ in the present experiments indicates a close relation between a charge gap and a spin gap in a novel phase transition in this system.

In summary, we have studied the magnetization of CeOs$_2$Al$_{10}$ up to 55 T for $H\parallel a$.  
The concave $M$-$H$ curve is observed not only below $T_0$ but in a paramagnetic region below $T_{\rm max}$ where $\chi_a$ and $\chi_c$ exhibit a maximum. 
We proposed that a Kondo singlet accompanied with a spin gap is formed below $T_{\rm max}$ and coexists with the AFM order below $T_0$. 
As the origin of the AFM order with  $M_{\rm AF}\parallel c$ at $H$=0, we propose the larger suppression of the spin degrees of freedom along the $a$-axis than along the $c$-axis.  

From very recent inelastic neutron scattering experiments by P. Deen $et\ al$.\cite{Deen}, we note that the spin gap in CeOs$_2$Al$_{10}$ is clearly open at $T$ = 35 K. 
This observation rules out a simple magnon gap below $T_0$ as a possible explanation for the observed spectral response in the inelastic neutron scattering experiments, thus implying that it should result from a spin singlet formation.

\end{document}